\documentclass[aps,prl,twocolumn,groupedaddress,floatfix]{revtex4}
\usepackage{graphicx}
\usepackage{amssymb}
\newcommand{\degc}{$^{\circ}$C }
\begin{document}

\title{Persistent holes in a fluid}
\author{Florian S. Merkt}
\author{Robert D. Deegan}
\author{Daniel I. Goldman}
\author{Erin C. Rericha}
\author{Harry L. Swinney}
\affiliation{Center for Nonlinear Dynamics,
                The University of Texas at Austin,
                Austin, TX 78712}
\date{\today}

\begin{abstract}

We observe stable holes in a vertically oscillated 0.5~cm deep
aqueous suspension of cornstarch for accelerations $a$ above
$10g$. Holes appear only if a finite perturbation is applied to
the layer.  Holes are circular and approximately 0.5~cm wide, and
can persist for more than $10^5$ cycles.  Above $a\simeq 17g$ the
rim of the hole becomes unstable producing finger-like protrusions
or hole division.  At higher acceleration, the hole delocalizes,
growing to cover the entire surface with erratic undulations. We
find similar behavior in an aqueous suspension of glass
microspheres.
\end{abstract}


\maketitle

The free surface of a fluid at rest in a container is flat.
Departures from flatness induce a restoring flow whether the fluid
is Newtonian, viscoelastic, or liquid crystalline: poke the
surface, and the resulting indentation will be filled by the
ensuing flow. In contrast, we have discovered that a vibrated
aqueous suspension of cornstarch or glass microspheres can
permanently support holes and vertical finger-like protrusions.

The catalog of interface morphologies in accelerated fluids is
broad and well documented. Sinusoidal acceleration produces
Faraday waves~\cite{faraday:1831},
solitons~\cite{lioubashevski:1996,lioubashevski:1999}, and
jets~\cite{Longuet-Higgins:1983}. Impulsive acceleration produces
the Richtmyer-Meshkov
instability~\cite{richtmyer:1960,meshkov:1969}, and continuous
acceleration the Rayleigh-Taylor~\cite{taylor:1950} instability,
which are characterized by spires and bubbles~\cite{alon:1995}.
The holes and fingers in vibrated cornstarch are distinct from
these phenomena because they display neither the oscillation about
a flat state of Faraday waves and solitons, the finite lifetime of
jets, nor the unbounded growth of the Richtmyer-Meshkov or
Rayleigh-Taylor instabilities.

\begin{figure}[hb!]
\begin{center}
\includegraphics[width=3.1 in]{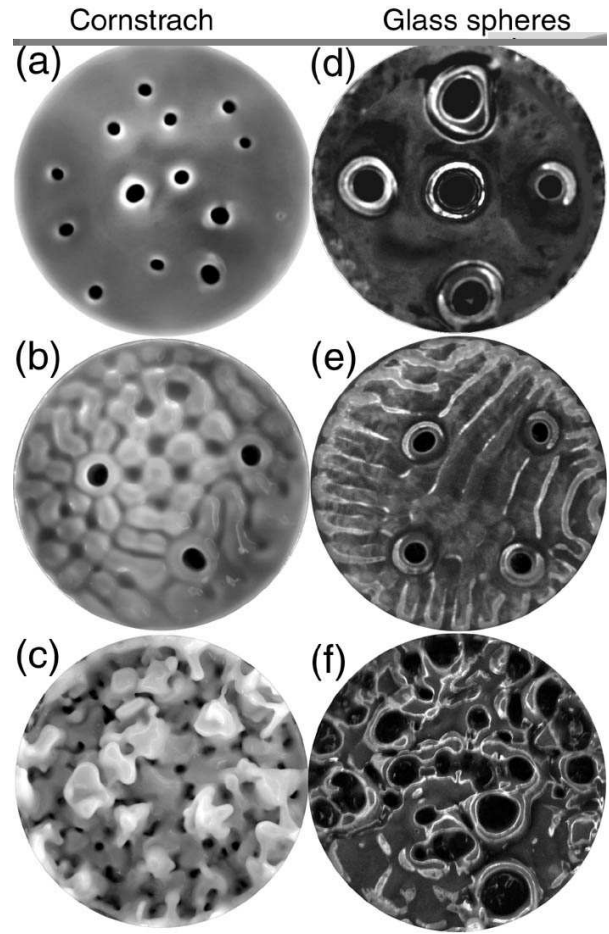}
\caption{\label{fig:comparison} Top view of  a vibrated layer of
(a-c) cornstarch and of (d-f) glass microspheres in liquids
matched in density to the particles.  Diameter is 9.4~cm. White
corresponds to the highest points, and black to depressions that
reach the container bottom. Holes without Faraday waves: (a) $a=
12g$, $f=150$~Hz; (d) $30g$, $100$~Hz. Holes with Faraday waves:
(b) $12g$, $60$~Hz; (e) $27.3g$, 92~Hz. Delocalized state: (c)
$30g$, $120$~Hz; (f) $30g$, $60$~Hz. }
\end{center}
\end{figure}

Figure~\ref{fig:comparison} shows holes in a vibrated aqueous
suspension of cornstarch~(a-b) and glass microspheres~(d-e).
Depending on the container acceleration and frequency, the initial
fluid surface is either flat or corrugated by Faraday waves. Above
a critical acceleration, a finite localized perturbation of the
fluid surface will grow into a stable cylindrical void that
extends from the top to the bottom of the fluid layer. The most
noteworthy feature of holes is their permanence: they do not close
despite the hydrostatic pressure of the surrounding fluid,
persisting as long as our observation ($>10^5$~cycles). At yet
higher accelerations, the holes lose stability to a temporally and
spatially erratic state (see Fig.~\ref{fig:comparison}(c) \& (f)).

\emph{Experiment.}  A layer of cornstarch or glass microspheres in
liquid was vertically oscillated sinusoidally with a frequency $f$
from 50~to~180~Hz and a peak acceleration $a$ up to 27$g$. The
fluid container was round with a 9.4~cm diameter, and made from an
aluminium base plate and acrylic sidewall. The container was
attached to the shaker through an insulating rod to avoid heat
transfer from the shaker to the container, and was sealed with an
acrylic top to reduce evaporation. The layer depth was 0.5~cm for
the cornstarch mixture and 0.2~cm for the glass microspheres. The
peak acceleration of the container was controlled to within $\pm
0.01g$.

The patterns were recorded with a 30~frames per second digital
camera and a 1000~frame per second video camera. Lighting was
provided by a ring of LEDs strobed at $f$ or $f/2$. The
illumination was such that peaks on the surface appear bright and
valleys dark.

We used unmodified, regular cornstarch consisting of 27\% amylose
and 73\% amylopectin from Sigma Aldrich. The powder was dried at
50\degc for a week and stored in a desiccator. A mixture was
prepared daily by combining 23.22~g of cornstarch, and 36.78~g of
a density-matched aqueous solution of CsCl with a density of
1.68~g/cm$^3$.  The viscosity of the suspension was measured in a
plate-plate geometry with a Paar Physica TEK150 rheometer. Though
all the quantitative results presented here are for cornstarch, we
also used a glass microsphere and water mixture for the
qualitative comparison in Fig.~\ref{fig:comparison}.  The glass
mixture was prepared with 29.37~g of $1-20~\mu$m glass balls from
Jaygo (Union, NJ), and 18.36~g of an aqueous solution of sodium
polytungstate with a density of 2.55~g/cm$^3$.

\begin{figure}[htp]
\begin{center}
\includegraphics[width=3in]{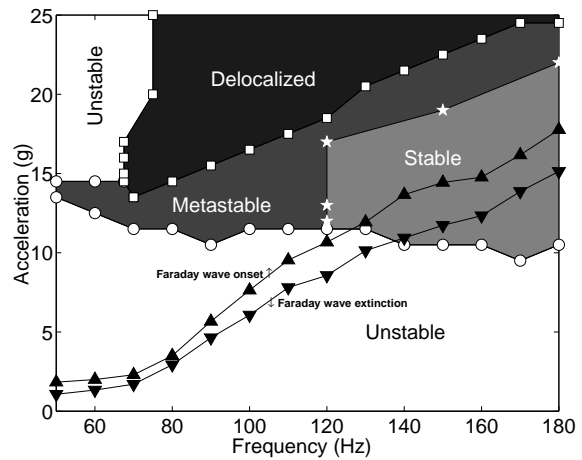}
\caption{\label{fig:phase} Phase diagram for a vibrated aqueous
cornstarch suspension as a function of acceleration and frequency.
For accelerations above the $\circ$, a surface perturbation will
form a hole.  In the metastable regime holes always collapse
within $10^5$ container oscillations; in the stable regime some
holes endure for at least $10^5$ oscillations. show the
acceleration threshold, below which holes collapse in less than
$10^4$~cycles. In the delocalized regime the state is chaotic as
shown in Fig.~\ref{fig:comparison}(c). Faraday waves appear with
increasing acceleration at $\blacktriangle$, and disappear with
decreasing acceleration at $\blacktriangledown$.}
\end{center}
\end{figure}

\begin{figure}[hb]
\begin{center}
\includegraphics[width=3in]{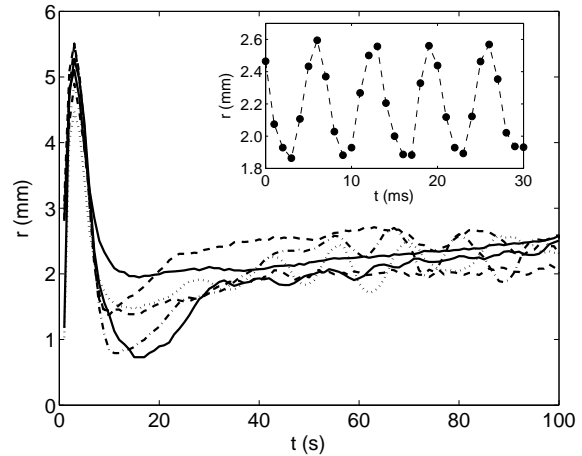}
\caption{\label{fig:size} Hole radius versus time in the
cornstarch suspension for 6 different trials at $15g$ and
$150$~Hz. At $t=0$ a jet of gas was applied to the surface for
2~s. The hole grows during this forcing, and immediately begins to
shrink after the gas is cutoff.  Inset: rapid oscillation of the
hole.  Note time scale.}
\end{center}
\end{figure}

\emph{Hole dynamics and size selection.}  The interface of a
vibrated fluid is flat at low accelerations and rippled by Faraday
waves above a critical acceleration~\cite{faraday:1831}. Holes
form in the cornstarch mixture only if an indentation deeper than
about 50\% the layer depth and 0.4-2.0~cm wide is applied to the
surface. We used a puff of compressed air delivered through a
nozzle pointed at the surface to form a crater. The subsequent
evolution of the crater depends on the parameters $a$ and $f$. For
all parameter values explored, the crater may close within
seconds. In the region of the phase diagram (Fig~\ref{fig:phase})
marked ``unstable'' all craters are annihilated by this rapid
decay. In regions marked ``stable'' and ``metastable'', the crater
may develop into a hole: the crater adjusts to a preferred size
(as shown in Fig~\ref{fig:size}) and remains fixed at that size
for an extended period of time.  At $f=120$~Hz the characteristic
diameter of the holes is 0.4~cm and at $f=180$~Hz 0.6~cm.  Holes
have a broad distribution of lifetimes, even for fixed $f$ and
$a$; in the metastable phase the maximum lifetime is less than
$10^5$~cycles and in the stable phase it is greater than
$10^5$~cycles.

Holes collapse via one of two mechanism depending on their
lifetime. Short lived holes collapse within a few seconds by a
uniform contraction. Longer lived holes with lifetimes of order
$10^4$ holes develop a hump on their rims, which then falls onto
the hole either covering it or rendering it so small that it
collapses rapidly by uniform contraction.  On occasion this latter
mechanism causes the hole to divide rather than collapse.

The cross-section of a hole is typically circular with a slight
eccentricity.  The shaft is smooth and slightly tapered towards
the base. At the base of the hole, there is usually a ribbon of
material about 0.1~cm high and wide that bisects the shaft.  These
features are not visible in Fig.~\ref{fig:comparison} due to the
lighting.   At accelerations immediately above the stability line
the material surrounding a hole remains level with the rest of the
layer.  At higher acceleration, the hole's rim rises about 0.05~cm
above the surrounding fluid.

\begin{figure}[hb]
\begin{center}
\includegraphics[width=3in]{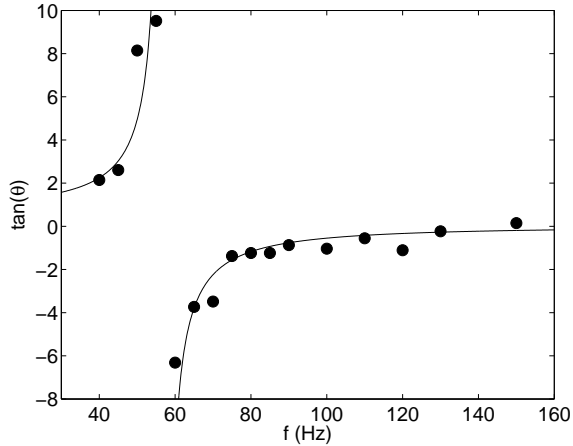}
\caption{\label{fig:phaselag} The tangent of the phase lag versus
frequency at $a=15g$  The solid line is a fit to
$\alpha/(1-(f/f_o)^2)$ where $\alpha$ and $f_o$ are fitting
parameters.}
\end{center}
\end{figure}

In addition to a slow meandering over hundreds of oscillations,
hole size oscillates synchronously with the driving frequency, as
shown in the inset of Fig.~\ref{fig:size}. The radius of a hole
evolves as $r(t)=\overline{r}+\delta r \sin{(2\pi f t + \theta)}$,
where $\overline{r}$ is the mean radius, $\delta r$ is the
amplitude of the oscillation, and $\theta$ is the phase relative
to the container motion $z(t)=z_o\sin{(2\pi f t)}$ where
$z_o=-a/(2\pi f)^2$ and $z$ is positive when the container is
above its rest point. $\delta r$ is typically $0.15 \overline{r}$
and also varies by about 20\% on a timescale of order $10^3f^{-1}$
. The phase lag is shown as a function of frequency in
Fig.~\ref{fig:phaselag}

Holes do not interact when their centers are separated by more
than two diameters. Therefore, as shown in Figure
\ref{fig:comparison}(a), holes do not form regular patterns.
Furthermore, they can be located anywhere in the container. Holes
do occasionally come sufficiently close to interact, and do so by
merging or one causing the other to collapse.

\emph{Faraday waves.} Though Faraday waves are excited in
cornstarch suspensions, the formation of holes is unrelated to
these waves, as can be seen from the existence of holes when no
Faraday waves are present, as in Fig.1(a) and (d). Furthermore,
the phase boundaries for holes and Faraday waves are distinct. The
Faraday transition is hysteretic and, moreover, the flat and
surface wave state can coexist.  As acceleration is increased at a
fixed frequency, small patches of surface waves appear.  The onset
of these are indicated by the $\blacktriangle$ in
Fig.~\ref{fig:phase}.  As the acceleration is raised further, the
patches grow and ultimately engulf the entire surface at values
typical 50\% higher than onset.  Upon decreasing the acceleration,
the surface waves breakup into patches which finally disappear at
accelerations indicated by the $\blacktriangledown$ in
Fig.~\ref{fig:phase}. Despite the ambiguity, there is a clear
difference in the parameter dependence for the onset of holes and
the onset of Faraday waves.

\begin{figure}[h!tb]
\begin{center}
\includegraphics[width=3in]{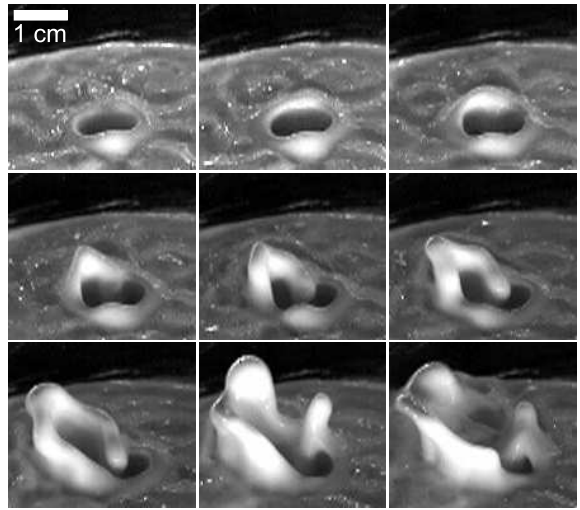}
\caption{\label{fig:deloc} Side view of the first steps of a
delocalizing hole in cornstarch. These photographs were taken
every 0.9~s.  Time increases from left to right and top to bottom.
An initial hump on the rim begins growing upward, it reaches a
maximum height, and then topples outward enlarging the area of
fluid motion. This process is repeated until the entire surface of
the liquid is active in the creation and destruction of vertical
structures and voids (see Fig.~\ref{fig:comparison}(c)).  These
images were taken at $a=$25$g$ and $f=80$~Hz. }
\end{center}
\end{figure}

\emph{Delocalization.} In the ``delocalized'' region in
Fig.~\ref{fig:phase}, a perturbation will generate a hole which
immediately grows a protrusion from its rim.  The sequence of
photos in Fig.~\ref{fig:deloc} shows the growth of such a
structure.  They are surprisingly tall and long-lived, reaching
heights of 2~cm and remaining upright for thousands of
oscillations. The protuberance eventually falls, nucleating a new
hole, and the growth of a finger is repeated. After many
repetitions the entire surface writhes with fingers and holes
yielding the spatially and temporally erratic state shown in
Fig.~\ref{fig:comparison}(c) and (f).

\begin{figure}[h!tb]
\begin{center}
\includegraphics[width=3in]{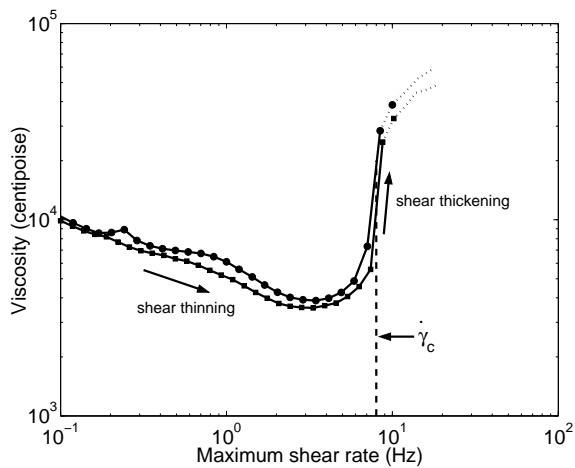}
\caption{\label{fig:visc} Apparent viscosity  versus maximum shear
rate $\dot{\gamma}_{max}$ for two identical samples of an aqueous
cornstarch suspension. The torque $\tau$ on the plate was measured
using a plate-plate geometry with a gap spacing $h$ of 0.10~cm and
a plate radius $R$ of 2.50~cm. The apparent viscosity was
calculated as it would for a Newtonian fluid: $\eta=2\tau/(\pi
\dot{\gamma}_{max} R^3)$. Since the shear rate increases with the
radius, our measurement of the viscosity is a convolution of the
geometry and the response of the fluid. The dotted line represents
measured values that are not reliable due to an instability in the
flow.  The cornstarch samples were the same as in
Fig.~\ref{fig:comparison} and the temperature was kept at $27.6\pm
0.2~^\circ$C.}
\end{center}
\end{figure}

\emph{Shear Thickening.} We attribute the formation of holes to
shear thickening, i.e., an increase of viscosity with shear rate.
A plot of the viscosity for two identically prepared cornstarch
suspensions is shown in Fig.~\ref{fig:visc}.  As is typical for
shear thickening fluid~\cite{barnes:1989}, an initial decrease of
the viscosity for increasing shear rate is followed by a rapid
increase at a critical shear rate, which in this case is
$\dot{\gamma}_c=8$~s$^{-1}$. Our conclusion is predicated on four
observations: the critical shear rate is similar in magnitude to
the shear rate at the throat of a hole; the phase lag of the holes
radius relative to the driving force indicates the material
response is primarily viscous; holes are unstable in a non-shear
thickening fluid; and holes are stable in shear thickening fluids
other than cornstarch suspensions.

The size of the hole varies synchronously with the driving
frequency.  The root-mean-squared shear rate at throat of the hole
is approximately the interface speed, $({2 \pi}/{\sqrt{2}}) \delta
r f$, divided by a length scale of order the depth of the layer
$h$. Using $\delta r/\overline{r} \approxeq 0.15$, $f=100$~Hz, and
$\overline{r}/h \approxeq 0.4$ yields a shear rate $\dot{\gamma}
\approxeq 27$~s$^{-1}$.  The similarity of this value to the
critical shear rate strongly suggests that shear thickening is an
essential ingredient for holes.

The frequency dependence of the phase lag $\theta$ indicates a
combined viscous and elastic response of the fluid.  Modelling the
hole motion as a spring in parallel with a dashpot (i.e., a Voigt
element, see \cite{ferry:1961} for example) yields
$\tan{\theta}=\alpha/(1-(f/f_o)^2)$, where $\alpha$ is a constant
proportional to the dissipation and $f_o$ is the resonant
frequency.  As shown in Fig.~\ref{fig:phaselag}, the data are well
modelled by this equation;the resonance frequency is around 58~Hz.
Since the Voigt element is primarily viscous at high frequency,
the phase data indicates that at frequencies above 60~Hz the
material response becomes dominantly viscous;  it is noteworthy
that stable holes only form in this high frequency region.

We also tried to form holes in silicon oil (a Newtonian fluid) and
various polybutadienes with viscosities 8.3, 28, and 72 Poise. In
all cases, however, the initial crater was backfilled within a few
hundred container oscillations.

In Barnes' review of shear thickening fluids, he  asserts that any
sufficiently concentrated suspension of solids in a fluid will
shear thicken~\cite{barnes:1989}. If our hypothesis that shear
thickening is crucial for hole formation is correct, then any
sufficiently dense suspension of solids ought to support
persistent holes.  For a dense suspension of glass microspheres
this is indeed the case, as shown in Fig.~\ref{fig:comparison}(d)
\& (e).

\emph{Conclusions.} We have shown that a vertically vibrated
aqueous suspension of cornstarch displays a number of unexpected
patterns--holes, fingers, and a delocalized state--that can only
be attained by the application of a finite perturbation.   Holes
have a preferred size, vibrate synchronously with the container,
and are unrelated to the well-known Faraday wave instability.  At
higher acceleration the hole becomes unstable to the creation of
finger-like structures, which at even higher accelerations are
responsible for the transition to a  delocalized state in which
the free surface is covered with fingers and holes.  Our
observation suggest that shear thickening is the key ingredient
for hole formation. The outstanding questions are how do holes
stay open, how can the finger like protrusions grow and remain
unright, and what drives the transition to the delocalized state?

\begin{acknowledgments}
We thank M.~Schr\"oter, W.D.~McCormick, P.~Heil, A.~Lee, W.~Zhang
for valuable suggestions, and P. Green and A. Jamieson for
assistance with the viscosity measurements. We acknowledge support
by the Engineering Research Program of the Office of Basic Energy
Sciences of the U.S. Department of Energy under grant number
DE-FG03-93ER14312.  FSM acknowledges support by the Friedrich
Ebert Stiftung.
\end{acknowledgments}

\end{document}